\documentclass[journal]{IEEEtran}
\usepackage{datetime}

\usepackage{bm}
\usepackage{cite}
\usepackage{graphicx}
\usepackage{amsmath}

\usepackage{CJK}
\usepackage{amsmath}

\usepackage{citesort}
\usepackage{amsfonts}
\usepackage{subfigure}
\usepackage{algorithmic}
\usepackage{algorithm}
\usepackage{amsmath,amssymb,verbatim,cite,citesort,amsopn,graphicx,citesort,url,color}
\usepackage{algorithm}
\usepackage{algorithmic}
\usepackage{multicol}
\usepackage{epsfig}
\usepackage{epstopdf}
\usepackage{balance}
\usepackage{stfloats}
\usepackage{color}  
\usepackage{subfigure}

\begin{document}

\begin{small}\title{Covert Communications with Constrained \\Age of Information }\end{small}

\author{{Yida Wang, Shihao Yan, Weiwei Yang, Yueming Cai}\\
\vspace{-0.15cm}
  \vspace{-1mm}
}

\maketitle

\begin{abstract}
In this letter, we consider the requirement of information freshness in covert communications for the first time. With artificial noise (AN) generated from a full-duplex (FD) receiver, we formulate a covertness maximization problem under the average age of information (AoI) constraint to optimize the transmit probability of information signal. In particular, the transmit probability not only represents the generation rate of information signal but also represents the prior probability of the alternative hypothesis in covert communications, which builds up a bridge between information freshness and communication covertness. Our analysis shows that the best transmit probability is not always $0.5$, which differs from the equal prior probabilities assumption in most
related works on covert communications. Furthermore, the limitation of average AoI enlarges the transmit probability at the cost of the covertness reduction and leads to a positive lower bound on the information transmit power for non-zero covertness.
\end{abstract}
\vspace{-2mm}
\begin{IEEEkeywords}
Covert communications, age of information, artificial noise, transmit probability.

\end{IEEEkeywords}

\section{Introduction} \label{sec1}

As an emerging security solution in wireless communications, covert communication is immune to the detection from a warden and thus can enable communication even when it is prohibited \cite{Yan Survey}. In general, the performance of covert communications depends on the warden's ability to distinguish between the noise and the information-carrying signal corrupted by noise \cite{Bash Square Root}. In the literature, \cite{Uninformed Jammer} introduced the artificial noise (AN) generated from a jammer to expand the uncertainty of the warden. Furthermore, \cite{FD conference,Shufeng,FD UAV} investigated the effect of AN transmitted by a full-duplex (FD) receiver in covert communications, which could be eliminated at the receiver by interference cancellation techniques.

The aforementioned works all focused on the throughput in communications with a constraint on covertness, while overlooked information freshness in communications. However, information freshness is of fundamental importance in many covert communications application scenarios, e.g., business activities, telehealth monitoring, or military actions, etc. Even if the covertness is guaranteed, the sluggish information update at the wireless receiver will still lead to severe threat to the property even life in related scenarios. It is noted that improving first-order metrics such as, providing high throughput or low delay, is inadequate when it comes to improving information freshness. Recently, a new metric termed as Age of Information (AoI) is proposed to characterize information freshness at the receiver \cite{AoI first}. It is defined as the time elapsed since the last successful decoded information containing the status update \cite{AoI book}. In this context, a natural and fundamental question that arises is \textit{``how to achieve communication covertness with the requirement of information freshness measured by AoI?''}. To the best of our knowledge, this question and similar issues have not been studied and this letter serves as the first attempt to address communication covertness subject to information freshness requirements.

In this letter, we consider a covert communication system in block Rayleigh fading channels, where a FD receiver always transmits AN to shield the potential information transmission with status updates. For the first time, we propose a framework of covert communications with the requirement of information freshness, i.e., adjusting the transmit probability of information signal with status updates to achieve the most communication covertness subject to the average AoI constraint. Specifically, from the aspect of the receiver, we derive the closed-form expressions for the average AoI and the expected detection error probability of the warden. Furthermore, we optimize the transmit probability in order to maximize the expected detection error probability subject to a given required average service time, which should be always larger than the average AoI. Our analysis shows that the strict limitation of the average AoI forces the transmitter to transmit information signal with a larger probability, even larger than $0.5$, which is different from the equal prior probabilities assumption in existing works (e.g., \cite{Uninformed Jammer,FD conference,Shufeng,FD UAV}). Moreover, the constrained average AoI also leads to a positive lower bound on the transmit power for the information signal at the transmitter to achieve non-zero covertness, which does not exist in the case without the constraint on information freshness.
\vspace{-2mm}
\section{System Model} \label{sec2}

\begin{figure}
\begin{center}
  \vspace{-1mm}
  \includegraphics[width=3.6in]{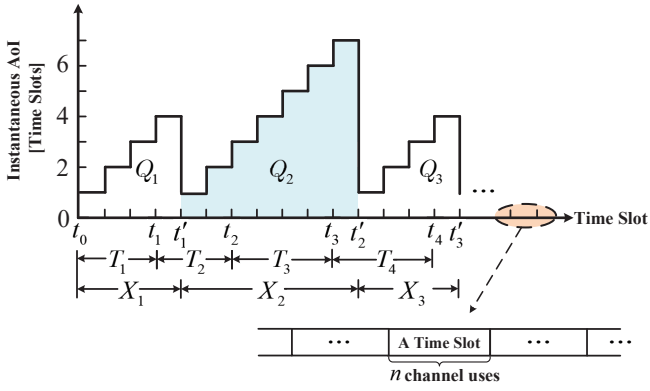}\\
  \vspace{-2mm}
\caption{One possible evolution of the instantaneous AoI versus the index of time slot for the considered covert communication system.}  \vspace{-5mm}\label{Fig2}
\end{center}
\vspace{-2mm}
\end{figure}

We consider a covert wireless communication system, where a transmitter (Alice) needs to send the sensitive information signal to a FD receiver (Bob) covertly under the supervision of a warden (Willie). Each of Alice and Willie is equipped with a single antenna, while, besides a single receiving antenna, Bob uses an additional antenna to transmit AN to confuse Willie. We denote the transmit power of Alice and Bob as $P_A$ and $P_B$, respectively.
Time is divided into slots and the time slot $t$ can only take discrete values, i.e., $t\!=\!0,1,2,\cdots$. It is noted that a time slot contains $n$ channel uses as shown in Fig. \ref{Fig2}.
And we consider the channels within the system follow block Rayleigh fading. It means that the channel gain within each time slot remains unchanged but varies independently from one time slot to another. The channels from Alice to Bob, Alice to Willie, and Bob to Willie, are denoted by $h_x$ and the mean values of ${\left| {{h_x}} \right|^2}$ are denoted by ${\lambda _x}$, where the subscript $x$ can be $AB$, $AW$ and $BW$, respectively. In addition, the self-interference channel of Bob is denoted by $h_{BB}$.

The randomized stationary transmission policy is considered in the system \cite{Chen He secure}. Specifically, at the beginning of each time slot, Alice transmits the sensitive information signal containing a new status update with a fixed probability $p$. It is noted that the transmit probability of information signal can be changed by switching on/off the transmission across time slots in practical.
When Alice transmits in a time slot, the signal received at Bob by employing interference cancellation techniques within this time slot is given by
\begin{equation}\label{ref001}
{\mathbf{y}_B}(j){\rm{ = }}\sqrt {{P_A}} {h_{AB}}\mathbf{x}_A(j) + \sqrt {\phi {P_B}} {h_{BB}}\mathbf{x}_B(j)+ {\mathbf{n}_B}(j),
\end{equation}
where $j\!\!=\!\!1,2,\cdots,n$ is the index of each channel use, $\mathbf{x}_A(j)$ is the signal transmitted by Alice satisfying $\mathbb{E}{\rm{[}}{\mathbf{x}_A}(j)\mathbf{x}_A^\dag (j){\rm{]}} = 1$, $\phi$ is the interference cancellation coefficient \cite{FD conference}, ${\mathbf{x}_B}(j)$ is the AN signal transmitted by Bob satisfying $\mathbb{E}{\rm{[}}{\mathbf{x}_B}(j))\mathbf{x}_B^\dag (j){\rm{]}} = 1$, and ${\mathbf{n}_B}(j)$ is the additive white Gaussian noise (AWGN) at Bob with the variance of $\sigma _B^2$. As such, the successful information decoding probability at Bob is given by
 \begin{equation}\label{ref001}
q{\rm{ = \!Pr}}\left\{\! {{\gamma _B} \!>\! {2^R} \!-\! 1} \right\} \!\!=\!\! \frac{{{P_A}{\lambda _{BB}}}}{{{P_A}{\lambda _{BB}}{\rm{ + }}\beta \phi {P_B}{\lambda _{AB}}}}\!\exp\! \left( \!\!{ - {\lambda _{AB}}\frac{{\beta \sigma _B^2}}{{{P_A}}}} \!\!\right),
\end{equation}
where ${\gamma _B}\!\! =\!\! \frac{{{P_A}{{\left| {{h_{AB}}} \right|}^2}}}{{\phi {P_B}{{\left| {{h_{BB}}} \right|}^2} + \sigma _B^2}}$ is the signal-to-interference-plus-noise ratio (SINR) at Bob, $R$ is the predetermined transmission rate from Alice to Bob and $\beta  = {2^R} - 1$. It is noted that when Bob fails to decode the information signal, the information will be discarded and will not be retransmitted \cite{Chen He energy trade-off}.

\subsection{AoI Evolution at Bob} \label{sec2-1}
Let $t_i$ represent the instant of the $i$-th transmission with the status update at Alice and $t_0\!=\!0$. Due to the unsuccessful information decoding and the aforementioned discard pattern, the information with the status update generated at Alice is not always be successfully received by Bob. Thus, we denote ${t'_k}$ as the instant of the $k$-th successful information decoding at Bob and $t'_0\!=\!0$. Then, the instantaneous AoI at time slot $t$ is given by
\begin{equation}\label{ref001}
A\left( t \right){\rm{ = }}\;\;t - r\left( t \right),
\end{equation}
where $r\left( t \right){\rm{ = }}\max \left( {{t'_k}{\rm{|}}\;{t'_k}{\rm{ < }}t} \right)$.

For the ease of understanding the AoI evolution at Bob, we further introduce some variables to simplify the calculation of the AoI. We denote $N_t$ as the number of successful information decoding at Bob till $t$, given by
\begin{equation}
{N_t}{\rm{ = }}\max \left( {k{\rm{|}}\;{t'_k}{\rm{ < }}t} \right),
\end{equation}
and denote $M_k$ as the number of transmission with the status update at Alice till $t_k$, given by
\begin{equation}
{M_k}{\rm{ = }}\max \left( {i{\rm{|}}\;{t_i}{\rm{ < }}{t'_k}} \right).
\end{equation}
We further define $X_k$ as the $k$-th interval time of two consecutive successful information decoding at Bob, i.e.,
\begin{equation}
{X_k}{\rm{ = }}{t'_k} - {t'_{k - 1}},
\end{equation}
and $T_i$ is defined as the $i$-th interval time of two consecutive transmission with the status update at Alice, i.e.,
\begin{equation}
{T_i}{\rm{ = }}{t_i} - {t_{i - 1}}.
\end{equation}
It is noted that ${t'_k} - {t_{{M_k}}}{\rm{ = }}1$ due to the aforementioned discard pattern. Hence, $X_k$ can be rewritten as
\begin{equation}\label{eq77}
{X_k}{\rm{ = }}\sum\limits_{i{\rm{ = }}{M_{k - 1}+1}}^{{M_k}} {{T_i}}.
\end{equation}

\vspace{-5mm}

\subsection{Detection at Willie} \label{sec2-2}
Most works on covert communications with a FD receiver achieve the covertness by varying the transmit power of AN across different time slots. However, varying transmit power across time slots is costly and hard to achieve \cite{Shufeng}. Thus, we assume $P_A$ and $P_B$ are fixed, which are known to Willie. Meanwhile, we consider that Willie only knows the channel distribution information (CDI) of $h_{BW}$, while still knows the instantaneous channel state information (CSI) of $h_{AW}$ from a conservative point of view. It is noted that Alice and Bob agree upon the pre-distributed secret keys in choosing the time slots to transmit, which is unknown to Willie.

As such, Willie needs to make a binary detection based on the observation vector $\mathbf{y}_W$ within a time slot, in which Alice does not transmit in the null hypothesis $H_0$ but it does in the alternative hypothesis $H_1$. Specifically, the composite received signal at Willie for each channel use is given by
\begin{equation}
\mathbf{y}_W(j){\rm{= }}\left\{ \begin{array}{l}
\!\!\!\!\sqrt {{P_B}} {h_{BW}}{\mathbf{x}_B}(j) +\! {\mathbf{n}_W}(j),\;\;\;\;\;\;\;\;\;\;\;\;\;\;\;\;\;\;\;\;\;\;\;\;\;\;H_0,\\\\
\!\!\!\!\sqrt {{P_B}} {h_{AB}}{\mathbf{x}_B}(j) +\! \sqrt {{P_A}} {h_{AW}}{\mathbf{x}_A}\!\!\left( j \right) \!+\! {\mathbf{n}_W}(j){\rm{,}}\;H_1,
\end{array} \right.
\end{equation}
where ${\mathbf{n}_W}(j)$ is the AWGN at Willie with the variance of $\sigma _W^2$.











\section{Analysis of the Average Age of Information} \label{sec3}
In this section, we analyze the average AoI at Bob. To this end, we first present Lemma 1 for the subsequent derivation of the AoI performance.

\vspace{2mm}
\noindent
\textbf{Lemma 1}: The first-order and the second-order moments for the interval time of two consecutive successful information decoding at Bob, respectively, are given by
\begin{equation}
\mathbb{E}\left( {{X_k}} \right){\rm{ = }}\frac{1}{{pq }},\label{eq1}
\end{equation}
\begin{equation}
\mathbb{E}\left( {X_k^2} \right){\rm{ = }}\frac{{2 - pq }}{{{{\left( {pq } \right)}^2}}}.\label{eq2}
\end{equation}
\noindent

\begin{IEEEproof}
The detailed proof is presented in Appendix A.
\end{IEEEproof}

As $t \to \infty $, the average AoI is given by
\begin{equation}\label{eq098}
\overline A  \!=\! \mathop {\lim }\limits_{t \to \infty }  \frac{{{1}}}{t}\sum\limits_{k = 1}^{{N_t}} {{Q_k}} {\rm{ \!=\! \! }}\mathop {\lim }\limits_{t \to \infty } \frac{1}{{{N_t}}}\sum\limits_{k = 1}^{{N_t}} {{Q_k}}  \!\!\times\!\! \frac{{{N_t}}}{t}{\rm{ = }}\frac{\mathbb{E}{\left( {{Q_k}} \right)}}{\mathbb{E}{\left( {{X_k}} \right)}},
\end{equation}
where ${Q_k}$ denotes the polygon area corresponding to ${X_k}$ as shown in Fig.2, $\mathop {\lim }\limits_{t \to \infty } \frac{1}{N_t}\sum\limits_{k = 1}^{N_t} {{Q_k}} {\rm{ = }}\mathbb{E}\left( {{Q_k}} \right)$ and $\mathop {\lim }\limits_{t \to \infty } \frac{N_t}{t} = \frac{1}{\mathbb{E}{\left( {{X_k}} \right)}}$. Indeed, $\frac{1}{\mathbb{E}{\left( {{X_k}} \right)}}$ is the steady state update rate of information at Bob. Specifically, the area of ${Q_k}$  is given by
\begin{equation}
{Q_k}{\rm{ = 1 + }}\left( {{\rm{1 + 1}}} \right){\rm{ + }} \cdots {\rm{ + }}\left( {1{\rm{ + }}{X_k}} \right){\rm{ = }}\frac{{X_k^2{\rm{ + }}{X_k}}}{2}.
\vspace{-2mm}
\end{equation}
 Thus, the $\mathbb{E}{\left( {{X_k}} \right)}$ can be rewritten as
\begin{equation}\label{eq099}
\vspace{-2mm}
\mathbb{E}\left( {{Q_k}} \right){\rm{ = }}\frac{\mathbb{E}{\left( {X_k^2} \right){\rm{ + }}\mathbb{E}\left( {{X_k}} \right)}}{2}.
\end{equation}

\vspace{2mm}
 With the help of Lemma 1 and (\ref{eq099}), the average AoI can be derived from (\ref{eq098}) in the following theorem.

\vspace{2mm}
\noindent
\textbf{Theorem 1}: The average AoI at Bob for the considered covert communication system is given by
\begin{equation}\label{eq11}
\bar A{\rm{ = }}\frac{\mathbb{E}{\left( {X_k^2} \right)}}{{2\mathbb{E}\left( {{X_k}} \right)}}{\rm{ + }}\frac{1}{2}{\rm{ = }}\frac{{\rm{1}}}{{pq }}.
\end{equation}

\noindent

\section{Achieving Covert Communications} \label{sec4}

From the perspective of Willie, the detection performance is normally measured by the detection error probability \cite{test statistical book,Gaussian Signal}, which is defined as
\vspace{-2mm}
\begin{equation}\label{eq4}
\vspace{-1mm}
\xi  = P\left( {{H_0}} \right){P_{FA}} + P\left( {{H_1}} \right){P_{MD}},
\vspace{-1mm}
\end{equation}
where $P\left( {{H_1}} \right) = {p}$ and $P\left( {{H_0}} \right) = 1 - {p}$ is the probability that Alice transmits or not, respectively, ${P_{FA}} = \Pr \left( {{D_1}|{H_0}} \right)$ is the false alarm probability, ${P_{MD}} = \Pr \left( {{D_0}|{H_1}} \right)$ is the miss detection probability, and $D_1$ and $D_0$ are defined as the events that Willie makes a decision in the favor of Alice transmitting or not, respectively. The optimal test for Willie to minimize the detection error probability is the likelihood ratio test (LRT), which can be transformed to the test of the average power received in a time slot \cite{Uninformed Jammer}. It is given by
\begin{equation}\label{eq6}
{P_W}\mathop {{\rm{\gtrless}}}\limits_{{D_{\rm{2}}}}^{{D_{\rm{1}}}} \tau,
\end{equation}
where $P_W$ is the average power received at Willie in a time slot, and $\tau $ is Willie's detector threshold, which can be further optimized to minimize the detection error probability. Considering the case of $n \to \infty $, $P_W$ is given by
\begin{equation}\label{eq5}
{P_W}{\rm{ \!=\! }}\left\{ \begin{array}{l}
\!\!\!{P_B}{\left| {{h_{BW}}} \right|^2}{\rm{ \!+\! }}\sigma _W^2,\;\;\;\;\;\;\;\;\;\;\;\;\;\;\;\;\;\;\;\;H_0,\\\\
\!\!\!{P_B}{\left| {{h_{BW}}} \right|^2}{\rm{ \!+\! }}{P_A}{\left| {{h_{AW}}} \right|^2}{\rm{ \!+\! }}\sigma _W^2{\rm{,}}\;\;\;H_1.
\end{array} \right.
\end{equation}

\subsection{Optimal Detection Threshold} \label{sec3-1}

Following the fact that the optimal strategy for Willie is to employ a radiometer, we next determine the optimal setting of Willie's detector threshold.

\vspace{2mm}
\noindent
\textbf{Theorem 2}:Using a radiometer for detecting Alice-Bob transmission, the optimal value of threshold to minimize the detection error probability at Willie is given by
\begin{equation}\label{eq8}
{\tau ^*}{\rm{ = }}\left\{ \begin{array}{l}
{\left| {{h_{AW}}} \right|^2}{P_A}{\rm{ + }}\sigma _W^2,\;\;\;\;{\left| {{h_{AW}}} \right|^2} \ge {\rho _{\rm{0}}},\\\\
{\rm{ + }}\infty ,\;\;\;\;\;\;\;\;\;\;\;\;\;\;\;\;\;\;\;\;\;\;{\left| {{h_{AW}}} \right|^2} < {\rho _{\rm{0}}},
\end{array} \right.
\end{equation}
where ${\rho _0}{\rm{ = }}\frac{{{P_B}}}{{{\lambda _{BW}}{P_A}}}\ln \left( {\frac{{{\rm{1}} - p}}{p}} \right)$ is the threshold of detection channel quality, and the corresponding minimum detection error probability at Willie is given by
\begin{equation}\label{eq9}
{\xi ^*}{\rm{ = }}\left\{ \begin{array}{l}
\left( {1 - p } \right)\theta_0 ,{\left| {{h_{AW}}} \right|^2} \ge {\rho _{\rm{0}}},\\\\
p ,\;\;\;\;\;\;\;\;\;\;\;\;\;{\left| {{h_{AW}}} \right|^2} < {\rho _{\rm{0}}},
\end{array} \right.
\end{equation}
where $\theta_0 {\rm{ = }}\exp \left( { - \frac{{{\lambda _{BW}}{{\left| {{h_{AW}}} \right|}^2}{P_A}}}{{{P_B}}}} \right)$.

\vspace{2mm}
\noindent
\begin{IEEEproof}
The detailed proof is presented in Appendix B.
\end{IEEEproof}

\subsection{Expected Detection Error Probability} \label{sec3-2}
Since Alice and Bob are unaware of the instantaneous CSI related to Willie, they have to rely on the expected value of detection error probability to access the feasible covertness.

\vspace{1mm}
\noindent
\textbf{Theorem 3}: The expected detection error probability is
\begin{equation}\label{eq12}
\overline {{\xi ^*}} {\rm{ = }}\left\{ \begin{array}{l}
\left( {1 - p } \right){\theta _{\rm{1}}},\;\;\;\;\;\;\;\;\;\;\;\;\;\;\;\;\;\;\;\;\;\;p  \ge \frac{1}{2},\\\\
\left( {1 - p } \right){\theta _{\rm{1}}}\varphi  + p \left( {{\rm{1}} - \varphi } \right),\;p  < \frac{1}{2},
\end{array} \right.
\end{equation}
where ${\theta _{\rm{1}}}{\rm{ = }}\frac{{{\lambda _{AW}}{P_B}}}{{{\lambda _{AW}}{P_B} + {\lambda _{BW}}{P_A}}}$ and $\varphi {\rm{ = }}\exp \left( { - \frac{{\rm{1}}}{{{\rm{1}} - \theta_1 }}  \ln \left( {\frac{{{\rm{1}} - p}}{p}} \right) } \right)$.

\begin{IEEEproof}
We use $\overline {{\xi ^*}} $ to denote the expected detection error probability over all realizations of ${h_{AW}}$. It is noted that the threshold of detection channel quality ${\rho _0}{\rm{ = }}\frac{{{P_B}}}{{{\lambda _{BW}}{P_A}}}\ln \left( {\frac{{{\rm{1}} - p}}{p}} \right)$ is a decreasing function of $p$, and ${\rho _0} = 0$ when $p  = \frac{1}{2}$. Thus, we conclude that ${\rho _0} > 0$ for $p  < \frac{1}{2}$, and ${\rho _0} \le 0$ for $p  \ge \frac{1}{2}$. Then we analyze $\overline {{\xi ^*}} $ in above two cases, respectively.

\vspace{1mm}
\noindent
\textit{Case \uppercase\expandafter{\romannumeral1}}: $p  < \frac{1}{2}$

From (\ref{eq9}), the expected ${\xi ^*}$ is given by
\begin{equation}\label{eq10}
 \begin{split}
\overline {{\xi ^*}} &{\rm{ = }}\mathbb{E}\left( {{\xi ^*}{\rm{|}}\;{{\left| {{h_{AW}}} \right|}^2} \ge {\rho _{\rm{0}}}} \right)\Pr \left( {{{\left| {{h_{AW}}} \right|}^2} \ge {\rho _{\rm{0}}}} \right)
\\
&{\rm{ + }}\mathbb{E}\left( {{\xi ^*}{\rm{|}}\;{{\left| {{h_{AW}}} \right|}^2}{\rm{ < }}\;\;{\rho _{\rm{0}}}} \right)\Pr \left( {{{\left| {{h_{AW}}} \right|}^2}{\rm{ < }}\;\;{\rho _{\rm{0}}}} \right).
 \end{split}
 \vspace{-4mm}
\end{equation}

\noindent
\textit{Case \uppercase\expandafter{\romannumeral2}}: $p  \ge \frac{1}{2}$

In this case, ${\rho _{\rm{0}}} \le {\rm{0}}$, hence ${\left| {{h_{AW}}} \right|^2} < {\rho _{\rm{0}}}$ is impossible to occur. The expected ${\xi ^*} $ can be obtained by simplifying (\ref{eq10}) as follows
\begin{equation}
\overline {{\xi ^*}} {\rm{ = }}\left( {{\xi ^*}{\rm{|}}\;{{\left| {{h_{AW}}} \right|}^2} \ge {\rho _{\rm{0}}}} \right).
\vspace{-2mm}
\end{equation}

With the aid of conditional expectation \cite{conditional expectation}, (\ref{eq12}) is derived by combining above two cases.
\end{IEEEproof}

\vspace{1mm}
\noindent
\textbf{Remark 1}:
It is worth mentioning that $p$ is not only the transmit probability at Alice but also the prior probability of the alternative hypothesis $H_1$ at Willie, which builds up a natural connection between information freshness and communication covertness. Naturally, we are interested in that how does $p$ affect simultaneously both the average AoI $\bar {A}$ and the expected detection error probability $\overline {{\xi ^*}}$. Therefore, we will optimize $p$ to achieve a trade-off between the $\bar {A}$ and $\overline {{\xi ^*}}$.

\vspace{-2mm}
\subsection{Covert Communications Design} \label{sec3-3}

From the perspective of the covert communication pair, i.e., Alice and Bob, we formulate the expected detection error probability maximization problem under the required average service time to optimize the transmit probability at Alice. In addition, the required average service time should be always greater than average AoI to guarantee the function of the service in applications \cite{aoi constrained1,aoi constrained2}. Specifically, the optimization problem is given by
\begin{equation}
\begin{array}{l}\label{eq13}
\textbf{\rm{P1}}\;\;\mathop {\max }\limits_p  \;\;\overline {{\xi ^*}} \\
\;\;\;\;\;\;\;s.t.\;\;\;\;\bar A \le \delta \\
\;\;\;\;\;\;\;\;\;\;\;\;\;\;\;\;{\rm{0}} \le p  \le {\rm{1}},
\end{array}
\end{equation}
where $\delta$ is the required average service time.

\noindent
\textbf{Theorem 4}: The optimal transmit probability ${p ^{\rm{*}}}$ of the optimization problem (\ref{eq13}) is
\begin{figure}[H]
\begin{center}
  \includegraphics[width=2.2 in]{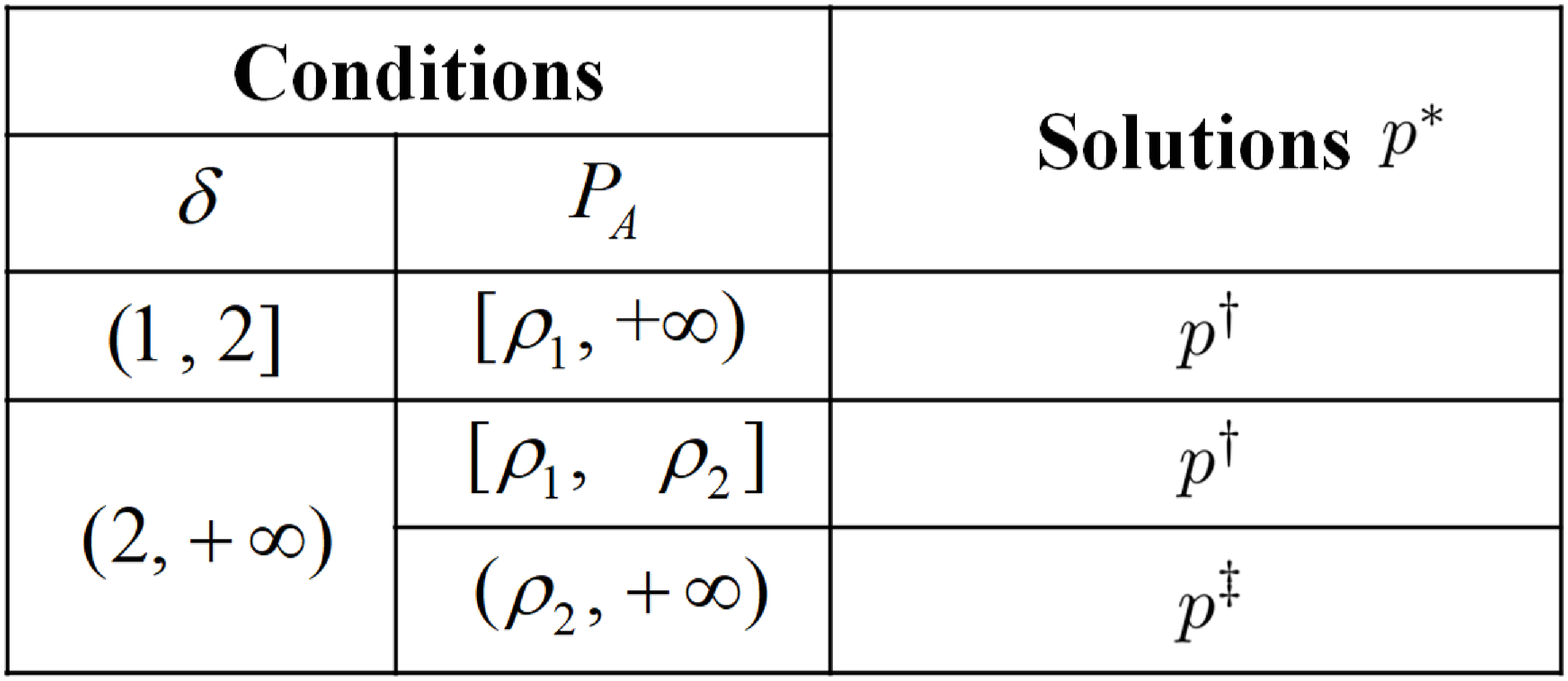}\\\vspace{-10pt}
\end{center}
\end{figure}
\noindent
where $\rho_1$ is the solution to $q\left( {{P_A}} \right){\rm{ = }}\frac{1}{\delta }$,  $\rho_2$ is the solution to $q\left( {{P_A}} \right){\rm{ = }}\frac{2}{\delta }$, ${p ^\dag }{\rm{ = }}\frac{1}{{q\delta }}$ and ${p ^\ddag }$ is given in (\ref{dadadada}).

\vspace{1mm}
\begin{IEEEproof}
We note that the limitation of average AoI can be rewritten as $p  \ge \frac{{\rm{1}}}{{q \delta }}{\rm{ > }}0$. Hence, there will be no feasible region of $p$ for $\frac{1}{{q\delta }} > 1$. Then, we can solve the optimization problem in two cases based on the construction of (\ref{eq12}), i.e., $\frac{1}{2} \le \frac{1}{{{q}\delta }} \le {\rm{1}}$ and $\frac{1}{{{q}\delta }}{\rm{ < }}\frac{1}{2}$, respectively.

\vspace{1mm}
\noindent
\textit{Case \uppercase\expandafter{\romannumeral1}}: $\frac{1}{2} \le \frac{1}{{{q}\delta }} \le {\rm{1}}$
\vspace{1mm}

The condition can be rewritten as $q \ge \frac{1}{\delta }$ for $\delta  \in {\rm{(1,}}\;{\rm{2]}}$ and $\frac{1}{\delta } \le {q} \le \frac{2}{\delta }$ for $\delta  \in (2,\;{\rm{ + }}\infty {\rm{)}}$. In this case, it is obvious that $\overline {{\xi ^*}} {\rm{ = }}\left( {1 - p } \right){\theta _{\rm{1}}}$ is a decreasing function of $p $. And the corresponding feasible region is $p  \in \left[ {\frac{{\rm{1}}}{{q\delta }}{\rm{,}}1} \right]$. Therefore, the optimal solution in this case is ${p ^\dag }{\rm{ = }}\frac{1}{{q\delta }}$.

\vspace{1mm}
\noindent
\textit{Case \uppercase\expandafter{\romannumeral2}}: $\frac{1}{{{q}\delta }}{\rm{ < }}\frac{1}{2}$
\vspace{1mm}

Similarly, the condition can also be written as $q{\rm{ > }}\frac{2}{\delta }$ for $\delta  \in \left( {2{\rm{,}}\;{\rm{ + }}\infty } \right)$. Due to the different constructions in (\ref{eq12}) for $\left( {\frac{1}{{pq }}{\rm{,}}\frac{1}{2}} \right)$ and for $\left[ {\frac{1}{2}{\rm{,}}1} \right]$, it is needed to compare the optimal value in these two cases. Based on the analysis in \textit{Case \uppercase\expandafter{\romannumeral1}}, the optimal value and optimal solution for $\left[ {\frac{1}{2}{\rm{,}}1} \right]$ is $\xi '{\rm{ = }}\frac{{{\theta _1}}}{2}$ and $p '{\rm{ = }}\frac{1}{2}$, respectively. And the optimal solution for $\left( {\frac{1}{{pq }}{\rm{,}}\frac{1}{2}} \right)$ can be obtained by a simple one-dimension searching, which is given by
\begin{equation}
p '' = \mathop {\arg \max }\limits_{1/q\delta  \le p  < 1/2} \;\left( {1 - p } \right){\theta _{\rm{1}}}\varphi  + p \left( {{\rm{1}} - \varphi } \right).
\end{equation}
The corresponding optimal value is $\xi '' = \left( {1 - p ''} \right){\theta _{\rm{1}}}\varphi \left( {p ''} \right) + p ''\left( {{\rm{1}} - \varphi \left( {p ''} \right)} \right)$. Therefore, the optimal solution in this case is given by
\begin{equation}\label{dadadada}
{p ^\ddag }{\rm{ = }}\left\{ \begin{array}{l}
\frac{1}{2}{\rm{,}}\;\;\;\;\;\;\;\;\;\;\;\;\;\xi ' \ge \xi '',\\
p ''{\rm{,}}\;\;\;\;\;\;\;\;\;\;\;\xi ' \;{\rm{ < }}\;\xi ''.
\end{array} \right.
\end{equation}

After some algebra calculations, we present the solution to the optimization problem (\ref{eq13}) in Theorem 4.
\end{IEEEproof}

\vspace{1mm}
\noindent
\textbf{Remark 2}: From Theorem 4, it is obvious that the optimal transmit probability in problem (\ref{eq13}) is not always $0.5$, which is opposite to the equal prior probabilities assumption in most related work on covert communications. This is caused by the constraint on the average AoI and the uncertainty characteristic of the fading channel $h_{BW}$. In addition, it is necessary to guarantee $\delta {\rm{ > }}1$ and ${P_A} \ge {\rho _{\rm{1}}}$ to achieve non-zero expected detection error probability, resulting from the average AoI constraint.

\section{Numerical Results} \label{sec4}

In this section, we present the numerical results and study the performance of the considered system. In our simulations, we set the predetermined transmission rate from Alice to Bob as $R\!=\!1$, the AWGN variances at Bob and Willie as $\sigma _B^{\rm{2}}{\rm{ = }}\sigma _W^{\rm{2}}{\rm{ = }}{\rm{-60}}$ dBm, the mean values of all fading channels as ${\lambda _{AB}} \!=\! {\lambda _{AW}} \!=\! {\lambda _{BW}} \!=\! {\lambda _{BB}}\!=\!1$, and the interference cancellation coefficient as $\phi=0.01$.

\begin{figure}
\centering  
\subfigure[]{
\label{Fig.sub.1}
\includegraphics[width=0.23\textwidth]{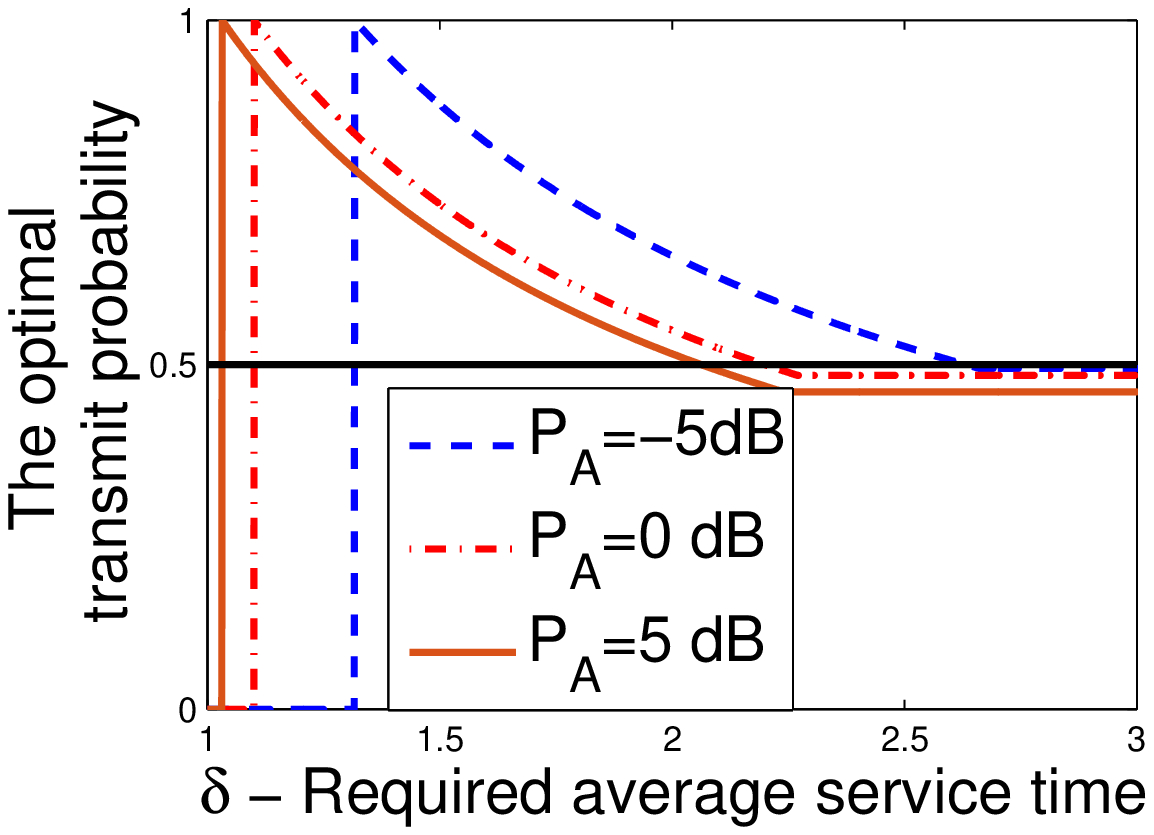}}
\subfigure[]{
\label{Fig.sub.2}
\includegraphics[width=0.23\textwidth]{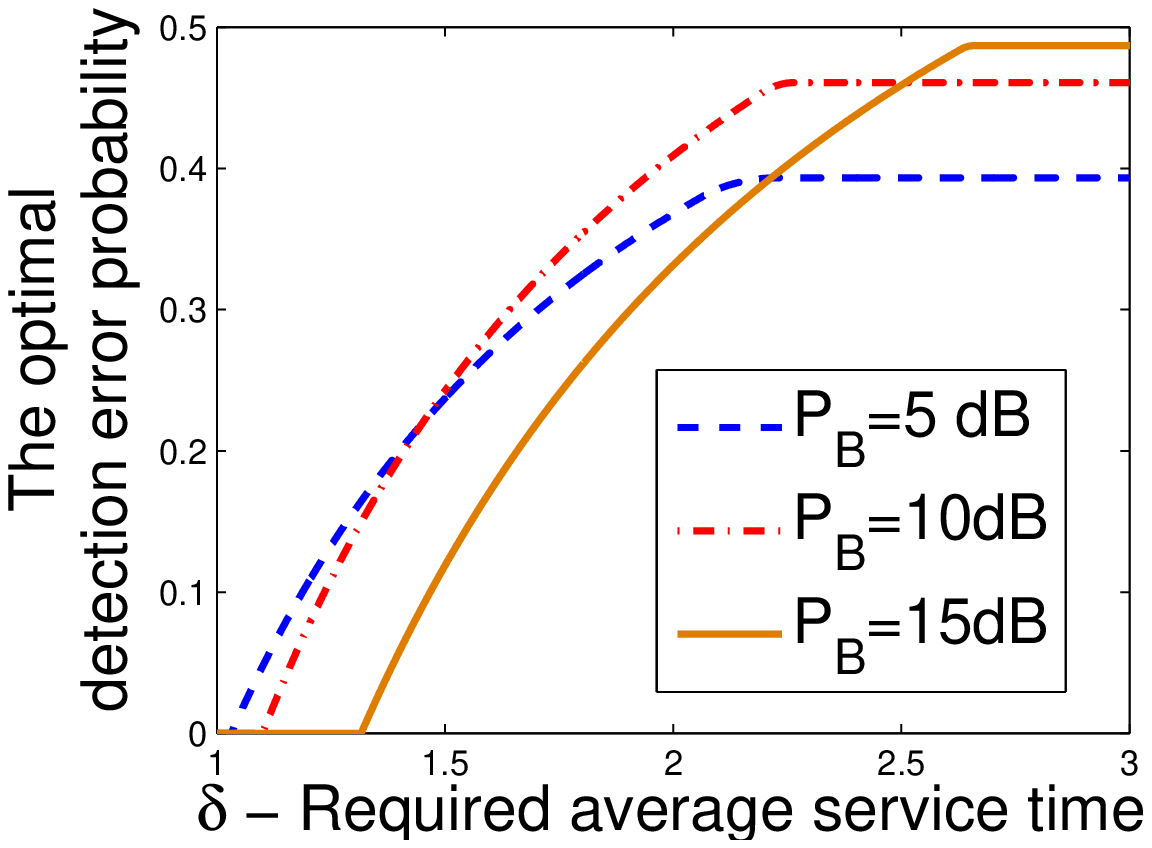}}
\caption{The required average service time versus: (a) the optimal transmit probability of information signal with $P_B=10$ dB; (b) the optimal expected detection error probability with $P_A=0$ dB}
\label{Fig.main}
\end{figure}

In Fig. \ref{Fig.sub.1}, we plot the optimal transmit probability $p^* $ versus the required average service time $\delta $ for different values of transmit power $P_A$. In this figure, we first observe that $p^* $ first decreases and then reaches a constant floor as $\delta $ increases. It is obvious that $p^* $ is not always $0.5$ as stated in Remark 2. We also observe that there may be no feasible solution to this optimization problem as $\delta$ decreases. This observation demonstrates that the positive lower bound on $P_A$ becomes stricter as $\delta$ decreases.

In Fig. \ref{Fig.sub.2}, we plot the optimal expected detection error probability $\overline {{\xi ^*}}$ versus the required average service time $\delta $ for different values of transmit power $P_B$. In this figure, we first observe that the optimal $\overline {{\xi ^*}}$ increases first and then remains unchanged as $\delta $ increases. It means relaxing $\delta$ properly is beneficial for improving $\overline {{\xi ^*}}$. Surprisingly, we also observe that a larger $P_B$ may lead to a lower $\overline {{\xi ^*}}$. It is caused by the fact that the the constraint on the constraint on AoI also becomes stricter as $P_B$ increases.

\section{Conclusion} \label{sec5}

In this letter, we proposed a framework for achieving covert communications with the requirement of information freshness, i.e., achieving the maximum expected detection error probability with constrained average AoI. Specifically, we optimized the transmit probability of information signal to obtain the trade-off between the covertness and the average AoI. Our analysis proved that transmit probability of $0.5$ is not always the best choice in the considered framework, which is different from the common equal prior probabilities assumption in covert communications. Meanwhile, the constraint on the average AoI forces the transmitter to transmit information signal with a higher probability and creates a positive lower bound on the information transmit power for achieving non-zero covertness.

\begin{appendices}
\section{Proof of Lemma 1}
The $k$-th interval time of two consecutive successful information decoding at Bob can be written as (\ref{eq77}). Thus, the expectation of ${X_k}$ can be rewritten by the law of total probability as follows
\begin{equation}
\begin{array}{ccccc}
\mathbb{E}\left( {{X_k}} \right) & {\rm{ \!\!\!\!=\!\! }}\sum\limits_{u{\rm{ = }}1}^\infty \!\! \mathbb{E}{\left(\! {\sum\limits_{i{\rm{ = }}{M_{k - 1}}{\rm{ + }}1}^{{M_k}} \!\!\!\!\!\!\!{{T_i}} {\rm{|}}\;{M_k} \!\!-\!\! {M_{k - 1}}{\rm{ = }}u} \!\right)} \mathbb{E}\left( {{M_k} \!\!-\!\! {M_{k - 1}}{\rm{= }}u} \right)\\
&\!\!\!\!\!\!\!\!\!\!\!\!\!\!\!\!\!\!\!\!\!\!\!\!\!\!\!\!\!\!\!\!\!\!\!\!\!\!\!\!\!\!\!\!\!\!\!\!\!\!\!\!\!\!\!\!\!\!\!\!\!\!\!\!\!\!\!\!\!{\rm{ = }}\sum\limits_{u{\rm{ = }}1}^\infty  {u\mathbb{E}\left( {{T_i}} \right)} {\left( {1 - q} \right)^{u - 1}}q.
\end{array}
\end{equation}

As the information generation at Alice follows a Bernoulli process, $T_i$ follows a geometric distribution with parameter $p$, leading to the fact $\mathbb{E}\left( {T_i} \right){\rm{ = }}\frac{{\rm{1}}}{p }$ and $\mathbb{E}\left( {T_i^2} \right){\rm{ = }}\frac{{2 - p }}{{{p ^2}}}$. Then we can derive (\ref{eq1}) and (\ref{eq2}) with help of {[16, Eq. 1.113]}.

\section{Proof of Theorem 2}
Based on (\ref{eq6}) and (\ref{eq5}), the false alarm probability and the miss detection probability are given by
\vspace{-2mm}
\begin{align}
 {P_{FA}}& = \left\{ \begin{array}{l}
{\rm{1,}}\;\;\;\;\;\;\;\;\;\;\;\;\;\;\;\;\tau  \le \sigma _W^{\rm{2},}\\
{\theta _{\rm{2}}},\;\;\;\;\;\;\;\;\;\;\;\;\;\;\tau \;{\rm{ > }}\;\sigma _W^{\rm{2}},
\end{array} \right.\\
{P_{MD}}&= \left\{ \begin{array}{l}
{\rm{0,}}\;\;\;\;\;\;\;\;\;\;\;\;\;\;\;\;\tau  \le \sigma _W^{\rm{2}}{\rm{ + }}{\left| {{h_{AW}}} \right|^2}{P_A},\\
1 - {\theta _3},\;\;\;\;\;\;\;\;\tau \;{\rm{ > }}\;\sigma _W^{\rm{2}}{\rm{ + }}{\left| {{h_{AW}}} \right|^2}{P_A},
\end{array} \right.
\end{align}
where ${\theta _{\rm{2}}}{\rm{\! =\! }}\exp \!\!\left( {\! - {\lambda _{BW}}\!\frac{{\tau  \!-\! \sigma _W^{\rm{2}}}}{{{P_B}}}} \!\!\right)$ and ${\theta _{\rm{3}}}{\rm{ \!=\! }}\exp \!\!\left( { \!- {\lambda _{BW}}\!\frac{{\tau  \!-\! \sigma _W^{\rm{2}} \!-\! {{\left| {{h_{AW}}} \right|}^2}}}{{{P_B}}}} \!\right)$.

\vspace{1mm}
\noindent
Then the detection error probability can be derived from (\ref{eq4}) as follows
\begin{equation}\label{eq7}
\begin{array}{ccccc}
\!\!\!\!\!\!\!\!\!\!\!\!\!\!\!\!\!\!\!\!\!\!\!\!\!\!\xi   {\rm{ = }}\left( {1 - p } \right)\Pr \left( {{P_W}{\rm{ > }}\tau {\rm{|}}\;{H_0}} \right){\rm{ + }}p \Pr \left( {{P_W}{\rm{ < }}\tau {\rm{|}}\;{H_1}} \right)\\
{\rm{ = }}\left\{ \begin{array}{l}
\!\!\!{\rm{1}} - p, \;\;\;\;\;\;\;\;\;\;\;\;\;\;\;\;\;\;\;\;\;\;\;\;\;\;\;\;\tau  \le \sigma _W^{\rm{2}},\\
\!\!\!\left( {{\rm{1}} - p } \right){\theta _{\rm{2}}},\;\;\;\;\;\;\;\;\;\;\;\;\;\;\;\;\;\;\;\;\;\sigma _W^{\rm{2}}{\rm{ < }}\tau  \le \sigma _W^{\rm{2}}{\rm{ + }}{\left| {{h_{AW}}} \right|^2}{P_A},\\
\!\!\!\left( {{\rm{1}} - p } \right){\theta _{\rm{2}}}{\rm{ + }}p \left( {{\rm{1}} - {\theta _{\rm{3}}}} \right){\rm{,}}\;\;\;\tau {\rm{ > }}\sigma _W^{\rm{2}}{\rm{ + }}{\left| {{h_{AW}}} \right|^2}{P_A}.
\end{array} \right.
\end{array}
\vspace{2mm}
\end{equation}

Then, we analyze the three possible cases in (\ref{eq7}) separately and obtain the optimal value of $\tau $ in (\ref{eq8}).

\vspace{1mm}
\noindent
\textit{Case \uppercase\expandafter{\romannumeral1}}: $\tau  \le \sigma _W^2$
\vspace{1mm}

Here, $\xi {\rm{ = }}1 - p $ cannot be minimized by $\tau $.

\vspace{1mm}
\noindent
\textit{Case \uppercase\expandafter{\romannumeral2}}: $\sigma _W^{\rm{2}}{\rm{ < }}\tau  \le \sigma _W^{\rm{2}}{\rm{ + }}{\left| {{h_{AW}}} \right|^2}{P_A}$
\vspace{1mm}

Here, $\xi $ is a decreasing function of $\tau $, hence Willie chooses the highest possible value of $\tau $, i.e., $\sigma _W^{\rm{2}}{\rm{ + }}{\left| {{h_{AW}}} \right|^2}{P_A}$.

\vspace{1mm}
\noindent
\textit{Case \uppercase\expandafter{\romannumeral3}}: $\tau {\rm{ > }}\sigma _W^{\rm{2}}{\rm{ + }}{\left| {{h_{AW}}} \right|^2}{P_A}$
\vspace{1mm}

In order to determine the optimal value of $\tau $ in this case, we need to analyze the first derivative of $\xi$, which is omitted here. And we come to the conclusion that if ${\left| {{h_{AW}}} \right|^2} \ge {\rho _{\rm{0}}}$, $\xi $ is a increasing function, otherwise $\xi $ is a decreasing function. It means that the optimal value of $\tau $ should be the lowest value or the highest value, respectively, i.e., $\sigma _W^{\rm{2}}{\rm{ + }}{\left| {{h_{AW}}} \right|^2}{P_A}$ or ${\rm{ + }}\infty $.

\end{appendices}

\end{document}